\documentclass[a4paper,10pt]{article}
\usepackage{graphicx,natbib}
\def\aap{A\&A\,  }
\def\aj{AJ  }
\def\apj{ApJ\,  }
\def\apjl{ApJ\,  }
\def\apss{Astrophysics and Space Science  }
\def\iaucirc{IAU circ.  } 
 
\def\mnras{MNRAS\,  }

\def\snr{SN\,1993J~}
\def\snrcinque{SN\,2005cf~}
\def\snrnove{SN\,1999ac~}
\def\snrel{SN\,2001el~}
\def\sn1987a{SN \,1987A\,}

\title
{
The physics of the optical light curve in  supernovae
}

\author{unknown}
\author{L. Zaninetti   \\
Dipartimento di Fisica \\
Via Pietro Giuria 1    \\
10125, Turin, Italy    \\
\footnote{zaninetti@ph.unito.it}\hspace{0.08cm}
{Corresponding author: zaninetti@ph.unito.it}}
\begin{document}
\maketitle

\begin{abstract}
We present a new formula which models
the rate of decline of supernovae (SN)
as given by the light curve in various bands.
The physical basis is the conversion
of the flux of kinetic energy into radiation.
The two main components of the model
are a power law dependence for the
radius--time relation and a decreasing
density with increasing distance
from the central point.
The new formula
is applied to
SN 1993J,
SN 2005cf, SN 1999ac,
and   SN 2001el
in different bands.

\end{abstract}

\section{Introduction}

The light curve (LC)  of supernovae  (SN) at a
given wavelength  $\lambda$
denotes the luminosity--time relation.
The astronomers work in terms of  apparent/absolute
magnitude and therefore the LC in SN is usually
presented as a magnitude versus time relation.
We have  two great  astronomical  classifications
for the LC: type I SN and type II SN.
The type I has a fast decrease in  magnitude
followed by a nearly linear increase.
In luminosity terms, the SN has a fast increase
followed by a nearly exponential decay.
The type II has a fast decrease in  magnitude
followed by oscillations, type IIb,
or a  plateau, type IIp; a decay  follows the plateau.
In this  complex morphology, we  will always specify
the type of SN under consideration.
The luminosity
is usually modeled by the  formula
\begin{equation}
L = L_{\lambda,0} \exp (- \frac{t}{\tau}) \quad ,
\end{equation}
where $L$ and $L_{\lambda,0}$ are the luminosity at time $t$ and
at $t=0$ respectively, and $\tau$ is the typical lifetime, see
\cite{deeming}.
As an example, the radioactive  isotope $^{56}$Ni
has  $\tau$ = 8.767 days. On introducing the apparent magnitude
$m_{\lambda}$, the previous formula becomes
\begin{equation}
m_{\lambda} = k^{\prime}_{\lambda}  +1.0857 (\frac{t}{\tau})
\quad ,
\label{mstandard}
\end{equation}
where $k^{\prime}_{\lambda}$  is a constant.
The absolute magnitude
$M_{\lambda}$  scales  in the same way:
\begin{equation}
M_{\lambda} = k^{\prime\prime}_{\lambda}  +1.0857 (\frac{t}{\tau}) \quad ,
\label{mgreatstandard}
\end{equation}
where $k^{\prime\prime}_{\lambda}$  is another
constant.
The observational fact  that, as an example,
in IC 4182 the LC has a half-life  of  56 days,
requires the production of
$^{56}$Co, see \cite{vanHise1974}.
The previous formula is an empirical  relation which is based
solely on observations rather than theory. The theory for SNII
LCs was first developed by \cite{Grasberg1971} and later
analytically and numerically explored by
\cite{Falk1973,Arnett1980,Arnett1989}. A model for the luminosity
in $H\alpha$ of supernovae as a function of time can be found  in
Figure 7 of  \cite{Chevalier1994}. The LCs of type Ia SN
have  been explained (including the secondary maximum) by a
time-dependent multigroup radiative transfer calculation, see
\cite{Kasen2006}. A  model  for  type II supernovae explosions
has been built including progenitor mass, explosion energy, and
radioactive nucleosynthesis, see \cite{Kasen2009}. The model
atmosphere code PHOENIX was used to calculate type Ia supernovae,
see \cite{Jack2011}. The previous works leave  a series of
questions unanswered or merely partially answered.
\begin{itemize}
\item
Given the observational fact that the
radius--time
relation in young SNRs follows a power law,
is it possible to
find a theoretical law of motion which fits the observations?
\item
Can a model of an expansion in the framework of the thin
layer approximation  produce the observed radius--time
relation?
\item
Can we express the flux of kinetic energy in the
framework of an approximate law of motion and a medium
characterized by a decreasing density?
\item Can we parametrize
the conversion of the flux of kinetic energy into  total observed
luminosity?
\item Can  we  parametrize  the fraction of
conversion of the total luminosity into the optical bands?
\end{itemize}
In order to answer these questions,
in Section \ref{motion} we analyze the
existing  equations of motion
for \snr   as well a new adjustable equation.
Section  \ref{syncro} reviews the basic formulas
of synchrotron emission and  reports
the conversion of flux of kinetic energy
into an observed band.
Section \ref{application} reports
the
application of the new formulas   to different SNs in
various bands.

\section{The equation of motion}

This  section  reviews three  existing parameters
for SNRs,
the power law model
and
a new solution in the framework  of the thin layer
approximation.

\subsection{Some existing solutions}
The  Sedov--Taylor  solution is
\begin{equation}
R(t)=
\left ({\frac {25}{4}}\,{\frac {{\it E}\,{t}^{2}}{\pi \,\rho}} \right )^{1/5}
\quad ,
\label{sedov}
\end{equation}
where $E$ is the energy injected into the process and $t$ is time,
see~\cite{Sedov1944,Taylor1950a,Taylor1950b,Sedov1959,Dalgarno1987}.
Our astrophysical  units are: time, ($t_1$), which is expressed in
years; $E_{51}$, the  energy in  $10^{51}$ erg; $n_0$, the
number density  expressed  in particles~$\mathrm{cm}^{-3}$~
(density~$\rho=n_0$m, where m = 1.4$m_{\mathrm {H}}$). In these
units, Equation~(\ref{sedov}) becomes
\begin{equation}
R(t) \approx  0.313\,\sqrt [5]{{\frac {{\it E_{51}}\,{{\it t_1}}^{2}}{{\it n_0}}}
}~{pc}
\quad .
\end{equation}
The Sedov--Taylor solution scales as
$R(t)\approx t^{2/5}= t^{0.4}$.

A second solution is connected with momentum conservation in the
presence of a constant density medium, see
\cite{Dyson1997,Padmanabhan_II_2001,Zaninetti2009a}. The
astrophysical radius in pc as a function of time is
\begin{equation}
R(t) =
\sqrt [4]{{{\it R_{0}}}^{3} \left(
 4.08\,10^{-6}\,{\it v_{1}}\, \left( t_1-t_{{0}} \right) +{\it R_{0}
} \right) } \, \mathrm{pc}
\quad ,
\end{equation}
where $t_1$ and $t_0$ are the  time in years, $R_0$ is the  radius in
pc  when  $t_1=t_0$, and $v_{1}$ is the velocity in
km s$^{-1}$ when $t_1=t_0$. The thin layer   solution in the
presence of a constant density medium scales as $t^{0.25}$. A
relativistic solution of  the thin layer approximation can be
found in \cite{Zaninetti2010e}.

A  sophisticated approach  as given by \cite{Chevalier1982a} and
\cite{Chevalier1982b} analyzes  self-similar solutions with
varying inverse power law exponents for the density profile of the
advancing matter, $R^{-n}$, and ambient  me\-dium, $R^{-s}$. The
previous assumptions give a law  of motion $R \propto
t^{\frac{n-3}{n-s} }$  when $n \, > 5$.

\subsection{The equation of motion as a power law }
\label{motion}
The equation of   the expansion  of an SNR
can  be modeled by a power law  of the type
\begin{equation}
R(t) = R_0 (\frac{t}{t_0})^{\alpha}
\label{rpower}
\quad ,
\end{equation}
where
$R$ is the radius  of the expansion,
$t$ is the time,
$R_0$ is the radius  at  $t=t_0$,
and  $\alpha$ is an exponent which  can be found from a
numerical analysis.
In order to find the unknown  parameters,
we analyzed the data
of supernova \snr, classified as
type IIb,
 which began to be visible
in M81 in 1993, see \cite{Ripero1993}, and presented a circular
symmetry for 4000 days, see \cite{Marcaide2009}. Its distance is
3.63~Mpc (the same as M81), see \cite{Freedman1994}.

\begin{figure}
\begin{center}
\includegraphics[width=6cm]{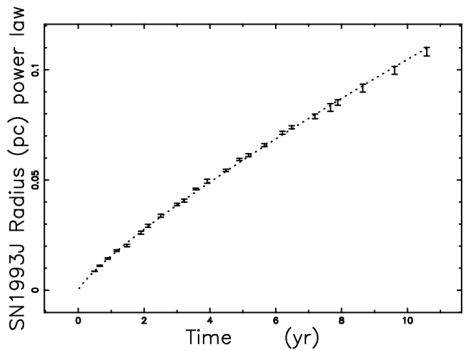}
\end {center}
\caption
{
The theoretical radius as given by the power law fit
represented by Equation~(\ref{rpower}) with $\alpha$ =0.828,
$R_0$=0.0087 pc and $t_0$ = 0.498 yr. The  astronomical data of
\snr  are represented by vertical error bars and are extracted
from Table 1 in Marcaide et al. (2009).
}
\label{1993pc_fit_power}
    \end{figure}

The velocity is
\begin{equation}
V(t) = \alpha R_0 (\frac{1}{t_0})^{\alpha} t^{(\alpha-1)}
\quad .
\label {vpower}
\end{equation}
As an example, Figure \ref{1993pc_fit_power}
reports the fit of
\snr.
We  have chosen  this SN
because:
\begin{itemize}
\item it presents a nearly spherical expansion,
\item the temporary  radius of expansion
has been measured for  $\approx$ 10 yr in
the radio band, see\cite{Marcaide2009}.
\end{itemize}
The observed   radius--time relation  of
\snr  allows us  to calibrate our model and
the application
of the least squares method
through the FORTRAN subroutine LFIT from
\cite{press} allows finding
$\alpha$ = 0.828 \,.
Therefore
the radius  is growing  more slowly
than a  free expansion with constant velocity,
$R\propto\,t$,
but   more quickly than the
Sedov--Taylor  solution, $R \propto  t^{0.4}$,
see  Equation (\ref{sedov}).

\subsection{An  adjustable equation of motion}

\label{sec_variable}
We assume that around the SNR the density of the
interstellar medium (ISM)
has
the following two piecewise dependencies
\begin{equation}
 \rho (R)  = \left\{ \begin{array}{ll}
            \rho_0                      & \mbox {if $R \leq R_0 $ } \\
            \rho_0 (\frac{R_0}{R})^d    & \mbox {if $R >    R_0 $ ~.}
            \end{array}
            \right.
\label{piecewiser}
\end{equation}
This assumption allows  us to set  up
the initial conditions, otherwise
the  dependence  $\rho (R) \propto (\frac{R_0}{R})^d$
will have a pole  at $R=0$.
At the moment of writing  there is not a clear
determination of the gradients
around the SN and therefore
$d$ can be considered a free parameter.

In this framework the SN is moving
and the density, which is at rest,
decreases as
an inverse power law with an exponent $d$
which can be fixed from the
observed temporal evolution of the radius.
The mass swept, $M_0$,
in the interval $0 \leq r  \leq R_0$
is
\begin{equation}
M_0 =
\frac{4}{3}\,\rho_{{0}}\pi \,{R_{{0}}}^{3}
\quad .
\end{equation}
The mass swept, $ M $,
in the interval $0 \leq r \leq R$
with $r \ge R_0$
is
\begin{equation}
M =
-4\,{r}^{3}\rho_{{0}}\pi \, \left( {\frac {R_{{0}}}{r}} \right) ^{d}
 \left(d -3 \right) ^{-1}
+4\,{\frac {\rho_{{0}}\pi \,{R_{{0}}}^{3}}{d-
3}}
+ \frac{4}{3}\,\rho_{{0}}\pi \,{R_{{0}}}^{3}
\quad .
\end{equation}
Momentum conservation in the thin layer approximation
requires  that
\begin{equation}
M v = M_0 v_0
\quad ,
\end {equation}
where  $v$   is  the velocity at $t$
and    $v_0$ is  the velocity at $t=t_0$.
The previous expression as a function of the radius
is
\begin{equation}
v  =
\frac
{
{{\it r_0}}^{3}{\it v_0}\, \left( 3-d \right)
}
{
3\,{{\it r_0}}^{d}{R}^{3-d}-{{\it r_0}}^{3}d
}
\quad .
\label{velclassic}
\end{equation}
In this differential equation of first order
in $R$, the
variables can be separated and an integration
term-by-term gives
the following
nonlinear equation ${\mathcal{F}}_{NL}$
\begin{eqnarray}
{\mathcal{F}}_{NL} =
 \left( 4{R_{{0}}}^{3}d-{R_{{0}}}^{3}{d}^{2} \right) R -3{R_{{0}}}^
{d}{R}^{4-d}+{R_{{0}}}^{4}{d}^{2}
+12{R_{{0}}}^{3}v_{{0}}t+3{R_{{0}
}}^{4}-4{R_{{0}}}^{4}d
\nonumber\\
+7{R_{{0}}}^{3}v_{{0}}d{\it t_0}+{R_{{0}}}^{3
}v_{{0}}{d}^{2}t
-7{R_{{0}}}^{3}v_{{0}}dt
-12{R_{{0}}}^{3}v_{{0}}{
\it t_0}-{R_{{0}}}^{3}v_{{0}}{d}^{2}{\it t_0}
=0
\quad  .
\label{nonlinear}
\end {eqnarray}
An approximate solution of
${\mathcal{F}}_{NL}(r) $
can be obtained
assuming that  \\
$3 R_0^d R^{4-d}$
$\gg$
$-(4 R_0^3 d-R_0^3 d^2)R$
\begin{equation}
 R(t) =
 ( {R_{{0}}}^{4-d}-\frac{1}{3}d{R_{{0}}}^{4-d}
( 4-d ) \nonumber \\
 + \frac{1}{3}
 ( 4-d ) v_{{0}}{R_{{0}}}^{3-d} ( 3-d )
 ( t-t_{{0}} )  ) ^{\frac{1}{4-d}}
\quad .
\label{asymptotic}
\end{equation}
Up to now, the physical units
have not been specified,
pc for length  and  yr for time
are perhaps acceptable choices.
With these units, the initial velocity $v_{{0}}$
is  expressed in pc yr$^{-1}$ and should be converted
into km s$^{-1}$; this means
that $v_{{0}} =1.02\,10^{-6} v_{{1}}$
where  $v_{{1}}$ is the initial velocity expressed in
km s$^{-1}$.

The astrophysical version of the above equation
in pc is
\begin{equation}
 R(t) =
 ( {R_{0}}^{4-d}-\frac{1}{3}d{R_{{0}}}^{4-d}
( 4-d )
 +3.402 \,10^{-7}
 ( 4-d ) v_{{1}}{R_{{0}}}^{3-d} ( 3-d )
 ( t_1-t_{{0}} )  ) ^{\frac{1}{4-d}} \, \mathrm{pc}
\quad ,
\label{radiusvarpc}
\end{equation}
where
$t_1$ and $t_0$ are  times in  years,
$R_0$ is the radius in pc  at $t_1=t_0$ and
$v_{1}$ is the velocity at
$t_1=t_0$
in km s$^{-1}$.
The approximate solution (\ref{asymptotic}) has the
following limit  as $t \to \infty$
\begin{equation}
 R(t) =C_{th} t^{\frac{1}{4-d}}
\quad  ,
\end{equation}
where
\begin{equation}
C_{th} =
 \left( \frac{1}{3}\,{\frac {{{\it R_0}}^{3}{\it v_0}\, \left( -3+d \right)
 \left( -4+d \right) }{{{\it R_0}}^{d}}} \right)
^{ \frac {1} {\left( 4-d \right)}}
\quad .
\end{equation}
On imposing
\begin{equation}
\alpha =  { \frac {1} {\left( 4-d \right)}}
\quad  ,
\end{equation}
we obtain
\begin{equation}
d   =
{\frac {4\,\alpha-1}{\alpha}}
\quad  .
\end{equation}
where  $\alpha$ is an observable  parameter defined in Section
\ref{motion}.
This means that the  unknown parameter $d$ can be
deduced from  the observed parameter $\alpha$.
More details  on
this model, as  well as a relativistic version, can be found in
\cite{Zaninetti2011a},
where conversely  the LC is not treated.

\section{The energy cascade }
\label{syncro} This  section contains the basic formula for the
synchrotron emission, the transformation of the mechanical flux
of energy into the observed luminosity, and the conversion of the
predicted flux at a given wavelength to the apparent magnitude.

\subsection{Synchrotron emission}

In SNR  we detect  non-thermal
emission
with intensity
\begin{equation}
I(\nu)
\propto
\nu^{+\beta}
\propto
\lambda^{-\beta}
\quad ,
\end{equation}
(where $\nu$  is the frequency,
      $\lambda$ the wavelength
and $\beta$ the power law index). As an example in the case of
\snr after the transition from optically thick to optically
medium, $\beta$  becomes $ \approx -0.6$ after  2500 days,  see
Figure 8 in \cite{MartiVidalb2011}. The conversion of the flux of
kinetic energy into synchrotron luminosity can be obtained by the
following  physical processes
\begin{itemize}
\item Turbulent evolution in the advancing shock in the framework
of  both the Kolmogorov and Kraichnan spectrum, see
\cite{Fan2010}. \item Particle acceleration in a turbulent
environment using a Monte Carlo approach for the diffusion and
acceleration of the particles, coupled to a magnetohydrodynamics
code in the SNR environment, see   \cite{Schure2010}. \item A
model for the  evolution  of the magnetic field in the advancing
layer, see \cite{Reynolds2011}. \item Diffusion of the
relativistic electrons from the position  of the advancing layer,
see Section 6  in \cite{Zaninetti2011a}.
\end{itemize}
The lifetime, $\tau_{syn}$,
for synchrotron  losses is
\begin{equation}
\tau_{syn} =
 39660\,{\frac {1}{H\sqrt {H\nu}}} \, \mathrm{yr}
\quad  ,
\end{equation}
where $H$ is the magnetic field in Gauss and   $\nu$ is  the
frequency of observation in Hertz, see \cite{lang}. The outlined
cascade  of physical processes can work if the following
inequalities  are verified
\begin{equation}
t_{cas}  < t_a <  \tau_{syn}
\quad  ,
\end{equation}
where $t_{cas}$  is the time scale  of
turbulence formation and $t_a$
is the time scale of electron acceleration.
In the following we will  assume that the
synchrotron emission is the main source
of luminosity. Two
radioactive decays will be considered in Section
\ref{application}.

\subsection{Non-thermal and thermal emission}

The synchrotron emission  in SNRs is detected  from $10^8$\ Hz of
radio-astronomy  to  $10^{19}$\ Hz of gamma astronomy which means 11
decades in frequency. At the same time, some  particular  effects,
such as absorption, transition  from optically thick  to optically
thin medium, line emission,   and the energy decay  of radioactive
isotopes ($^{56}$Ni, $^{56}$Co) can    produce  a change  in the
concavity of the flux  versus frequency relation, see  the
discussion about Cassiopea A in Section 3.3 of \cite{Eriksen2009}.
A comparison between  non-thermal and thermal emission (luminosity
and surface brightness distribution)  can be found  in
\cite{Petruk2007}, where it is possible to find some observational
tests which allow the estimation  of the parameters characterizing the
cosmic ray injection on supernova remnant shocks. At the same
time,  a technique  to isolate the synchrotron emission from the thermal
emission is  widely used, as an example see X-limb of  SN1006
\cite{Katsuda2010}.

\subsection{The temporal evolution}
The density of kinetic energy, $K$,
is
\begin{equation}
K = \frac{1}{2}\rho   V^2
\quad,
\end{equation}
where $\rho$ is the density and $V$ the velocity.
In presence of an area $A$ and when the velocity is  perpendicular
to that area,
the flux of kinetic energy
$L_m$ is
\begin{equation}
L_m = \frac{1}{2}\rho A  V^3
\quad,
\end{equation}
which in SI  is measured in J s$^{-1}$ and in CGS in erg $s^{-1}$
see formula (A28) in
\cite{deyoung}. In our  case, $A=4\pi R^2$, which means
\begin{equation}
L_m = \frac{1}{2}\rho 4\pi R^2 V^3
\quad ,
\end{equation}
where $R$  is the instantaneous radius of the SNR and
$\rho$  is the density in the advancing layer.
The source of synchrotron luminosity
is assumed here to be  proportional to
the flux of kinetic energy.
The density in the advancing  layer is  assumed
to scale as  $R^{-d}$,
which means that
\begin{equation}
L_m  \propto R ^{2-d}  V^3
\quad .
\end{equation}
This  last assumption  is connected  with  the
adjustable equation of motion
which is derived in a decreasing density environment,
see Section \ref{sec_variable}.
On adopting  this point of view,  $d$  is
an unknown  parameter
which allows matching theory  and observation.
The temporal and velocity evolution
are
given by the  power law
dependencies of Equations (\ref{rpower}) and
(\ref{vpower})
and therefore
\begin{equation}
L_m  \propto  t^{-\alpha d+5\alpha-3}
\label{kineticflux}
\quad  .
\end{equation}
The  synchrotron luminosity
$L_{\lambda}$
and  the observed  flux  $S_{\lambda}$
at a given wavelength    $\lambda$
are assumed   to be proportional
to the  mechanical  luminosity
and therefore
\begin{equation}
S_{\lambda} =  S_0 \,(\frac{\lambda}{\lambda_{obs}})^{-\beta}
 (\frac {t}{t_0})^{-\alpha d+5\alpha-3}
\quad ,
\label{fluxtime}
\end{equation}
where $S_0$ is the  flux
when   $t=t_0$
at a given wavelength
$\lambda_{obs}$.
The  apparent magnitude
at a given color  $c$,
where $c$ can be  $U,B,V,R$ or $I$,
is
\begin{equation}
m_{\mathrm c}  = k_c  - 2.5  \log_{10}
{\int Se_{\mathrm \lambda} I_{\lambda} d\lambda}
\quad,
\label{defmag}
\end {equation}
where  $Se_{\lambda}$ is the sensitivity function
in the region
specified by the wavelength  $\lambda$,
$k_c$  is a constant,
and  $I_{\lambda}$ is the energy flux
reaching the earth.
We now define a sensitivity function for a
pseudo-monochromatic
color   system
\begin {equation}
Se_{\lambda} = \delta (\lambda -\lambda_i)
\quad  i=U,B,V,R,I
\quad,
\end {equation}
where $\delta$ denotes the Dirac delta function, see
\cite{deeming}. In this system the apparent magnitude is
\begin{equation}
m_{\mathrm c}  = k_c  - 2.5  \log_{10} I_{\lambda}
\quad .
\label{defmagmono}
\end {equation}

On assuming  that the intensity of emission and
the flux of kinetic energy as given by (\ref{fluxtime})
are directly proportional,
we obtain
\begin{equation}
m_{\mathrm c}  =
- 2.5\,{\frac { \left( -\alpha\,d+5\,\alpha-3 \right) \ln  \left( t
 \right) }{\ln  \left( 2 \right) +\ln  \left( 5 \right) }}+k_c
\quad ,
\label{defmagnostra}
\end{equation}
where $k_c$  is a constant:
\begin{equation}
k_c  = - 2.5  \log_{10}
\bigl ( S_0 \,(\frac{\lambda}{\lambda_{obs}})^{-\beta}\bigr)
+k_b
\quad  ,
\end {equation}
and  $k_b$ is a constant.

In the previous  equations we  have three unknowns:
$\alpha$, $k_c$ and $d$.
In the case  of \snr   the value of $\alpha$ is deduced
from the data of the expansion.
On fixing two times in the observed LC,
$t=t_0$ and
$t=t_1$, we have two corresponding magnitudes
$m_0$ and  $m_1$.
The resulting nonlinear system  of two equations in
two unknowns can therefore be solved.

The  $(C_1-C_2)$ color can be expressed as
\begin{eqnarray}
(C_1-C_2)= \nonumber \\
m_{\mathrm 1} - m_{\mathrm 2}  =
k_{12}  - 2.5  \log_{10}
\frac
{\int Se_{\mathrm 2} I_{\lambda} d\lambda}
{\int Se_{\mathrm 1} I_{\lambda} d\lambda}
\quad,
\label{bv}
\end {eqnarray}
where
$k_{12}$ is a constant
and  $I_{\lambda}$ is the energy flux
reaching the earth.
In a pseudo-monochromatic  color system
\begin{equation}
C_1 - C_2 =
k_{12} - 2.5 \beta  \log_{10}  (\frac{\lambda_2}{\lambda_1})
\quad .
\end {equation}
According to the previous equation, the color of a SN
should  be constant with  time.
As an example in the case of \snrcinque  (type Ia)
(B--V) became stable after   the first 120 days \cite{Pastorello2007}.
The constancy of  the
color  has  been obtained with the assumption that the spectral
index is constant with time. The spectral index in the radio
varies  considerably  but  becomes constant, $\beta \approx  -
0.7$, after  $\approx$ 700 days, see Figure 8 in
\cite{MartiVidalb2011}. Late time photometric observations of \snr
show that $(i-R) =-0.1$ in the  interval 692 days  $< t <  $ 3260
days, further on
  $(e-i)= -0.26$  at 3245 days
and
 $(e-i)= -0.29$  at 3504 days
which means a small variation, $|\Delta (e-i)| =0.03$ in 259 days,
see  Table  3  in \cite{Zhang2004}.
In other words, the constancy
of the color can be applied after $\approx$ 700 days.

More precisely,
the observed luminosity at time $t$
can be expressed introducing
the initial  mechanical  luminosity,
$L_{m0}$, defined  as
\begin{equation}
L_{m0} = \frac{1}{2}\rho_0 4 \pi R_0^2 V_0^3
\quad ,
\end{equation}
where the index $0$  stands  for the first measurement.
The astrophysical  version of the above equation is
\begin{equation}
L_{m0} =
{1.39\times 10^{41}      }\,{\it n_1}\,{{\it R_1}}^{2}{{\it v_{10000}}}^{
3}
\mathrm{ergs}\,\mathrm{s}^{-1}
\quad ,
\label{kineticfluxastro}
\end{equation}
where $n_1$   is the  initial number density expressed
in units  of
particles cm$^{-3}$,
$R_1$  is  the initial radius expressed
in units of pc, and
$v_{10000}$ is the initial  velocity expressed in
units of 10,000 km s$^{-1}$.
The spectral luminosity, $ L_{\nu} $,
at a given frequency $\nu$
is
\begin{equation}
L_{\nu} =  4 \pi  D^2  S_{\nu}
\quad  ,
\end{equation}
with
\begin{equation}
S_{\nu} =  S_0  (\frac{\nu}{\nu_0})^{\beta}
\quad  ,
\end{equation}
where  $S_0$   is the flux
observed at  the frequency
$\nu_0$  and  $D$ is the  distance.
The total observed luminosity, $L_{tot} $,
is
\begin{equation}
L_{tot} =
\int_{\nu_{min}}^{\nu_{max}}  L_{\nu} d \nu
\quad  ,
\end{equation}
where
${\nu_{min}}$ and
${\nu_{max}}$
are the  minimum and maximum frequencies  observed.
The  total observed
luminosity
can  be expressed as
\begin{equation}
L_{tot} = \epsilon  L_{m0}
\label{luminosity}
\quad  ,
\end{equation}
where  $\epsilon$  is  a constant  of conversion
from  the mechanical luminosity   to  the
total observed luminosity in the synchrotron emission.
The fraction  of the total  luminosity deposited  in a
color  $f_c$  is
\begin{equation}
f_c  =
\frac
{
{{\it \nu_{c,min}}}^{\beta+1}-{{\it \nu_{c,max}}}^{\beta+1}
}
{
{{\it \nu_{min}}}^{\beta+1}-{{\it \nu_{max}}}^{\beta+1}
}
\quad  ,
\end{equation}
where  $\nu_{c,min}$  and  $\nu_{c,max}$
are the minimum and maximum frequency  of  a color.
Table \ref{tablecolors} presents some values of
$f_c$  for the most important optical bands.
\begin{table} [h!]
\label{tablecolors}
\begin{center}
      \caption
      {
      Table of the values of $f_c$ when
      $\nu_{min}= 10^7$\ Hz,
      $\nu_{max}= 10^{18}$\ Hz
      and  $\beta=-0.7$.
         }
         \begin{tabular}{crrr}
            \hline
           \hline
colour        & $\lambda$  ($\AA$) &  FWHM  ($\AA$) & $f_c$\\
            \hline
U            &  3650   &  700  & 6.86 $\times  10^{-3}$  \\
B            &  4400   &  1000 & 7.70 $\times  10^{-3}$   \\
V            &  5500   &  900  & 5.17 $\times  10^{-3}$  \\
$H\alpha$    &  6563   &  100  & 0.56 $\times  10^{-3}$  \\
R            &  7000   &  2200 & 9.32 $\times  10^{-3}$   \\
I            &  8800   &  2400 & 7.5  $\times  10^{-3}$   \\
            \hline
            \hline
         \end{tabular}
   \end{center}
   \end{table}
At the time of writing, the
number  density in  the advancing layer is  unknown
and  we can  therefore  define  $\epsilon n_1$
as  the constant which allows
adjusting  theory and observations.
About $\epsilon$ it should be said
that by definition $\epsilon < 1$.
The rapid rise in intensity in a SN can be modeled
by Equation (\ref{piecewise})
for  the radiative transfer
when a time dependent
transition from optically thick to optically thin medium
is considered.
The solution of the  radiative transfer equation
for the specific     intensity per unit frequency,
$I_{\nu}$,
at the end of an astrophysical  object,  is
\begin{equation}
I_{\nu}(\tau_{\nu})  =
I_{\nu}( 0  )  \exp {(-\tau_{\nu})}
+ G_{\nu} (1  -\exp {(-\tau_{\nu})}    )
\quad  ,
\end{equation}
where $\tau_{\nu} $ is the optical depth, $G_{\nu}$ is the source
function, and  $I_{\nu}( 0  )$ the intensity beyond the
astrophysical object, see equation (1.30) in \cite{rybicki}. On
considering only the intensity of the object ($I_{\nu}( 0  )=0$)
the previous formula becomes
\begin{equation}
I_{\nu}(\tau_{\nu})  =
G_{\nu} (1  -\exp {(-\tau_{\nu})}    )
\quad  ,
\label{transition}
\end{equation}
where $\tau_{\nu}$=1   represents  the value at which the intensity
is $63\%$ of the source function.
The temporal  transition from optically thick to optically thin medium  before
the maximum
can be modeled by imposing $ \tau_{\nu}= \frac{t}{t_a}$
where $t_a$ is a typical time.
This is an `ad hoc'  function  that allows  of modeling
the transition before and after  the maximum   and the consequent change  of concavity
of the  LC as function of time.
The time  $t_a$  can vary from the few seconds of
a  Gamma Ray Burst  (GRB)
to the few days  of the optical bands.

A  logarithmic form  of Equation (\ref{transition})
introduces the  apparent magnitude $m_{\nu}$
\begin{equation}
m_{\nu}=
- 1.085\,\ln  \left(  1- \,{{\rm e}^{- \,{\frac {t}{{
\it t_a}}}}} \right) +{\it k_{\nu} }
\label{equationmagnitude}
\quad ,
\end{equation}
where $k_{\nu}$ is a constant.

We are now ready to introduce  the
two  phase model which can be
characterized  by the following two
piecewise dependencies
\begin{equation}
m_{\nu}   = \left\{ \begin{array}{ll}
              - 1.085\,\ln  \left(  1- \,{{\rm e}^{- \,{\frac {t}{{
\it t_a}}}}} \right) +{\it k_{\nu} }
                     & \mbox {if $t \leq t_0 $ } \\
            - 2.5\,{\frac { \left( -\alpha\,d+5\,\alpha-3 \right) \ln  \left( t
 \right) }{\ln  \left( 2 \right) +\ln  \left( 5 \right) }}+k_c
     & \mbox {if $t >    t_0 $ ~.}
            \end{array}
            \right.
\label{piecewise}
\end{equation}

\section{Applications}
\label{application}

\begin{figure}
\begin{center}
\includegraphics[width=6cm]{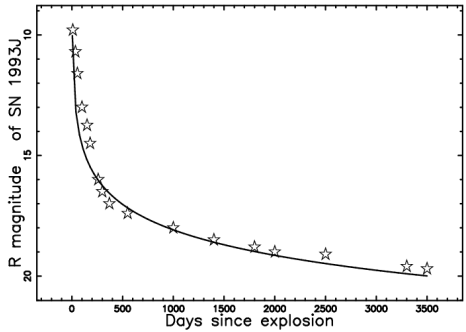}
\end {center}
\caption
{ The $R$ LC of \snr over 10 yr (empty stars)
and theoretical curve as given by (\ref{defmagnostra}) (full
line). In this case  $t_0$ = 5 days, $d$ = 3.075, $k_c$ = 7.543 and
$\alpha$ = 0.828. The data are extracted by the author from Figure 5
in Zhang et al. (2004).
} \label{1993magtime}
    \end{figure}

Figure \ref{1993magtime}
reports the decay of the $R$   magnitude of
\snr, which is type IIb, as well our theoretical curve.

\begin{figure}
\begin{center}
\includegraphics[width=6cm]{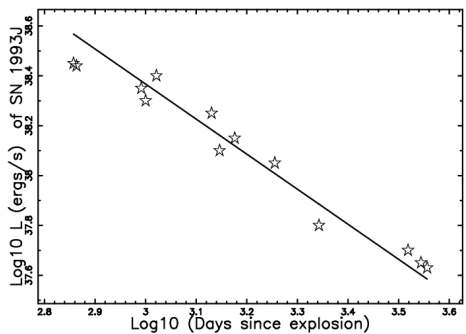}
\end {center}
\caption
{ The Log-Log {\it i}-band  (H$\alpha$ luminosity) LC of \snr over 10 yr (empty stars)  and theoretical curve as
given by (\ref{luminosity}) (full line). In this case  $t_0$ = 5
days, $d$ = 3.075, $\epsilon n_1=0.1$  and  $\alpha$ = 0.828. The
data are extracted by the author from Figure 7 in
 Zhang et al. (2004).
} \label{1993halfatime}
    \end{figure}

The theoretical temporal evolution of
the $H\alpha$ luminosity of
\snr as well  the data are  reported in
Figure~\ref{1993halfatime}.
The  $H\alpha$ luminosity which is derived from the  $i-$band   is
also fitted by the model of \cite{Chevalier1994}.

\begin{figure}
\begin{center}
\includegraphics[width=6cm]{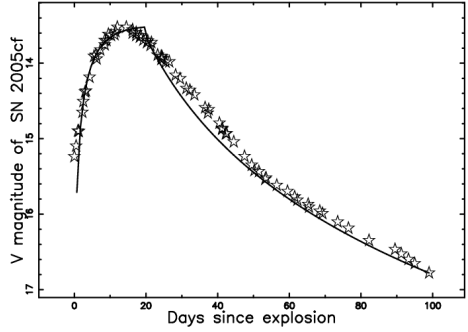}
\end {center}
\caption
{ The $V$ LC of \snrcinque (empty stars)  and
theoretical curve as given by the two phase model (\ref{piecewise})
(full line). The first phase  is modeled by $t_a =5$  days
 and
$k_{\nu}$ = 13.5  and the second phase  by $t_0$ = 20.5 days, $d$ =
3.547, $k_c$ = 7.810 and $\alpha$ = 0.828. The data are the final
S-corrected $V$ magnitudes reported in Table 6 of
Pastorello et al. (200).
}
\label{2005magvtutto}
    \end{figure}

\begin{figure}
\begin{center}
\includegraphics[width=6cm]{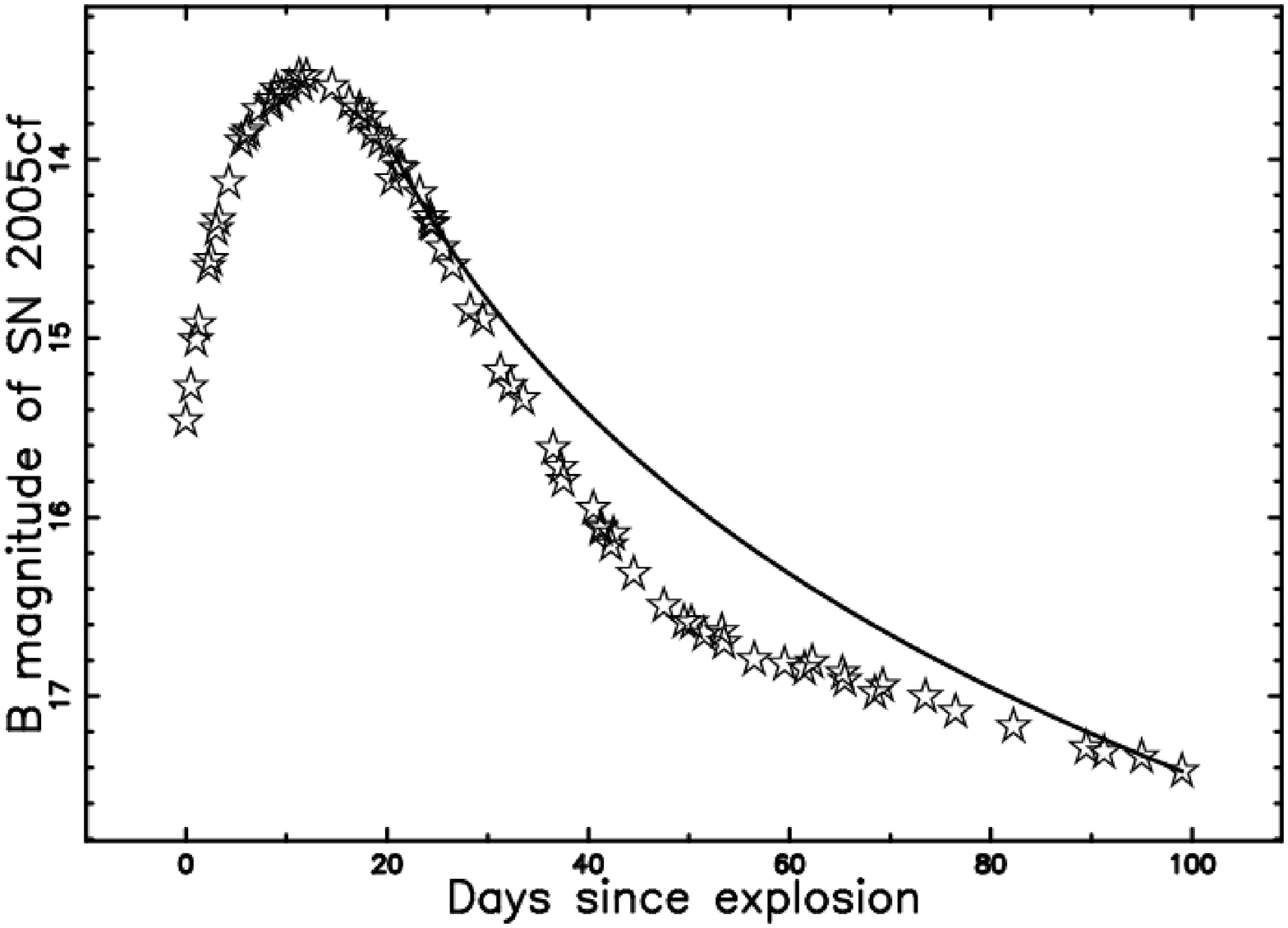}
\end {center}
\caption
{ The $B$ LC of \snrcinque (empty stars)  and
theoretical curve as given by (\ref{defmagnostra}) (full line). In
this case  $t_0$ = 20.5 days, $d$ = 3.83, $k_c$ = 7.286 and
$\alpha$ = 0.828. The data are the final S-corrected $V$ magnitudes
as reported in Table 6 of
Pastorello et al. (2007).
}
\label{2005magbtime}
    \end{figure}

\begin{figure}
\begin{center}
\includegraphics[width=6cm]{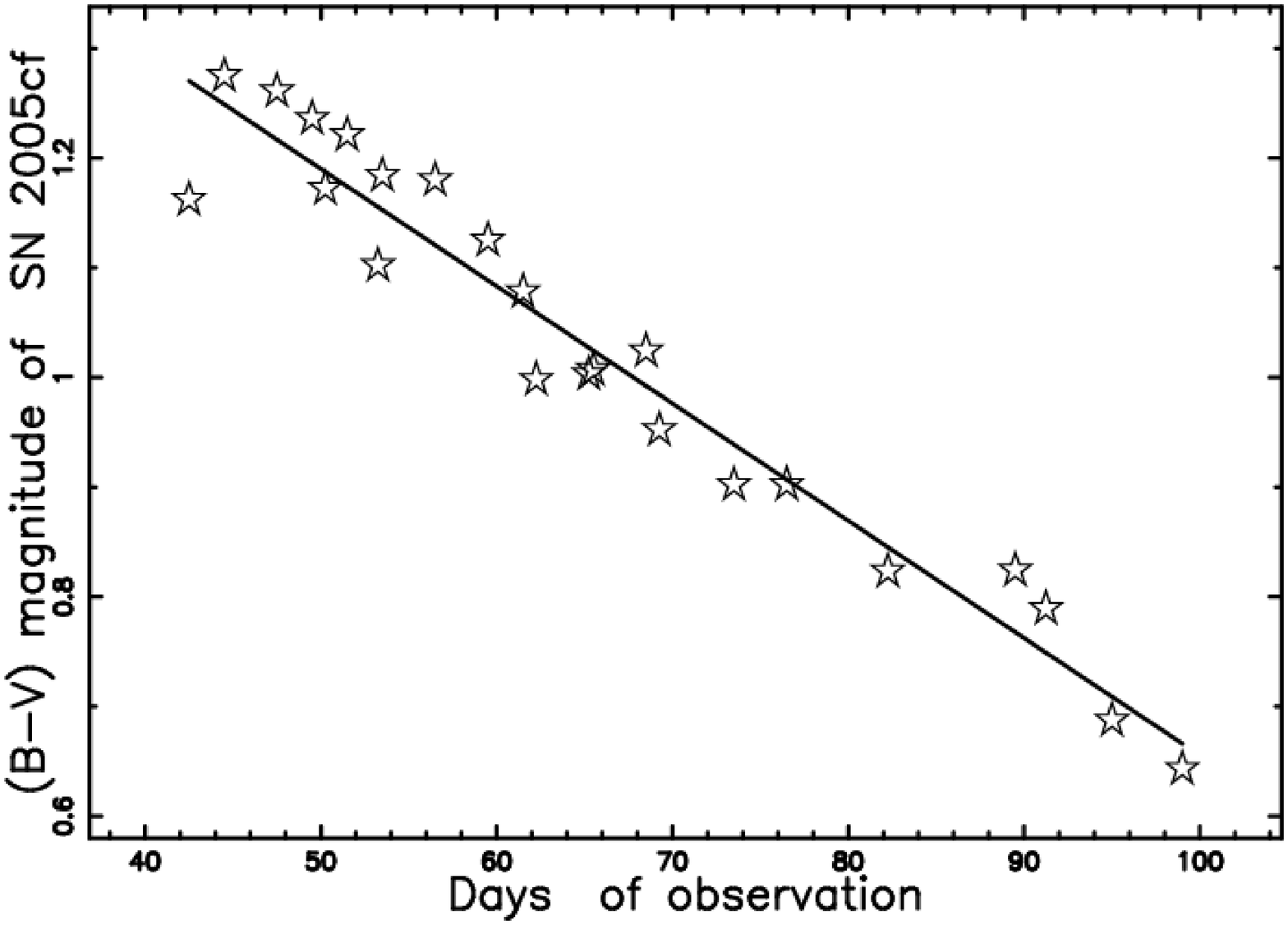}
\end {center}
\caption
{
The (B--V)  color evolution  of \snrcinque
(empty stars)  and the relative fitting
straight line (full line).
The time  ranges from   40 days to 100 days.
\label{2005magbvtime}
}
    \end{figure}

\begin{figure}
\begin{center}
\includegraphics[width=6cm]{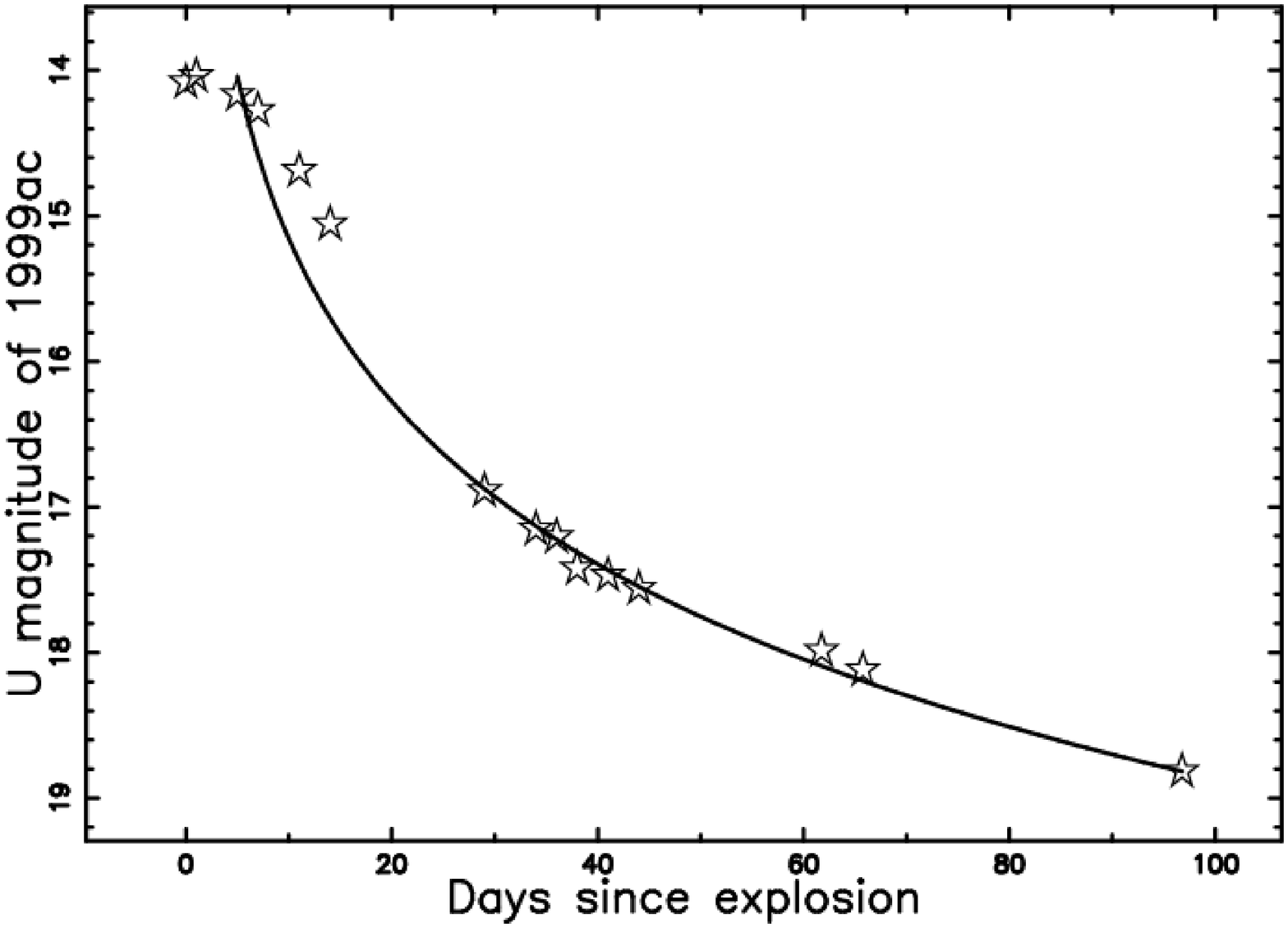}
\end {center}
\caption
{
The $U$ LC  of \snrnove
(empty stars)  and theoretical curve
as given by (\ref{defmagnostra}) (full line).
In this case  $t_0$ = 5 days,
$d$ = 3.17, $k_c$ = 11.44  and $\alpha$ = 0.828.
The data are those reported at CDS.}
\label{1999acmagutime}
    \end{figure}

\begin{figure}
\begin{center}
\includegraphics[width=6cm]{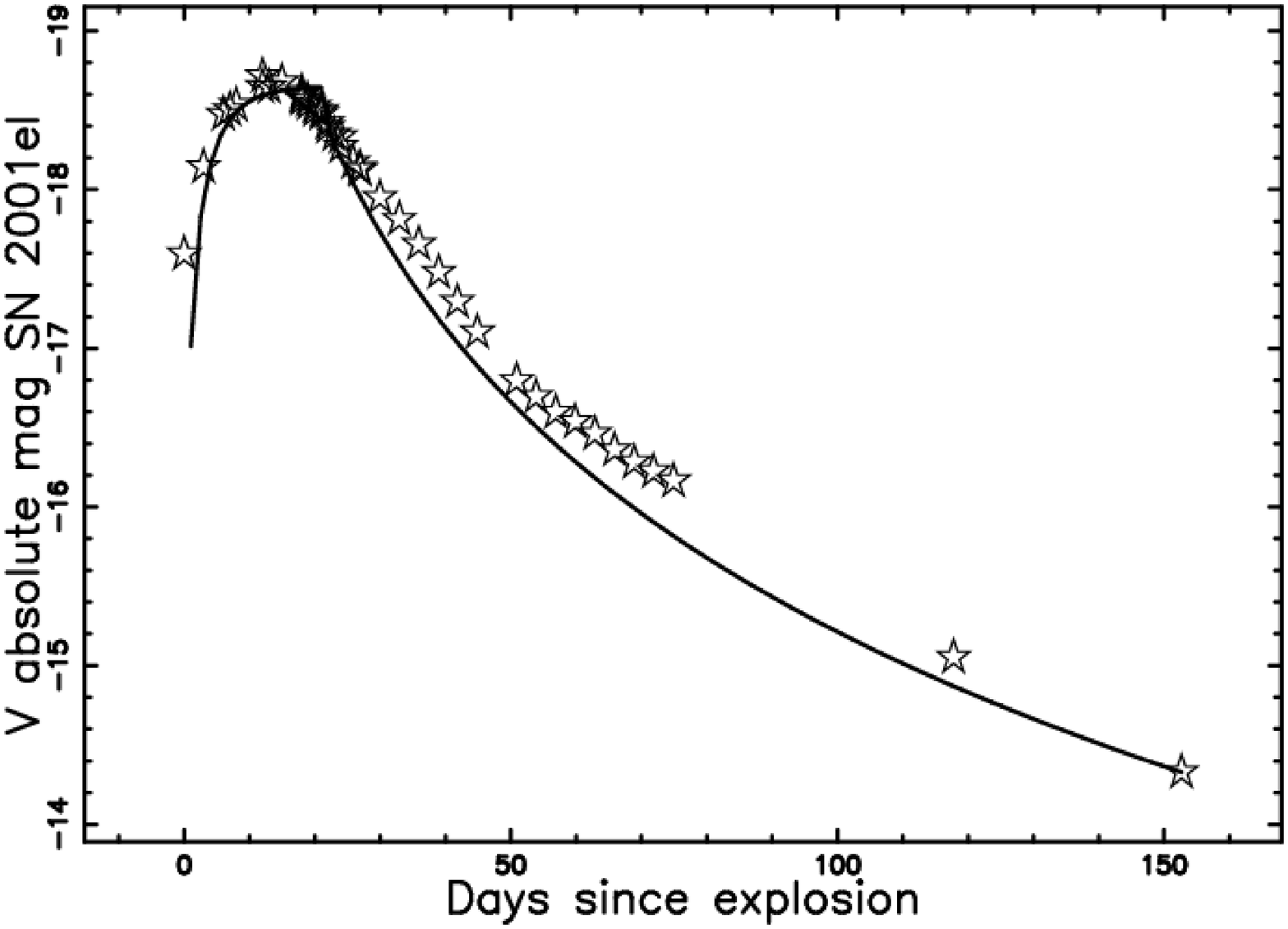}
\end {center}
\caption { The $V$
LC  of \snrel (empty stars)
in absolute magnitude
and
theoretical curve as given by the two phase model (\ref{piecewise})
(full line).
The first phase  is modeled by $t_a =4$  days
 and
$k_{\nu}=-18.64$,  and the second phase  by $t_0$ = 21 days, $d$ =
3.702, $k_c=-24.84$ and $\alpha$ = 0.828. The data are extracted
from  Table 3 of
Krisciunas et al. (2003)
and  the adopted distance
modulus  is  $\mu$ = 31.4 according to
Kasen et al. (2006).
}
\label{2001elmagvtutto}
    \end{figure}

A second  example  is   \snrcinque  (type Ia) which   has  been
analyzed  in \cite{Pastorello2007}; Figure \ref{2005magvtutto} and
Figure \ref{2005magbtime} report the decay of the $V$ and  $B$
magnitude of \snrcinque  as well our theoretical curve.
The (B--V)  color evolution  of \snrcinque
is reported in Figure~\ref{2005magbvtime}.
Only  the  second phase  is reported.
A third example is the sample of 44 type Ia supernovae which have
been observed in the UBVRI bands, see \cite{Jha2006}. We selected
\snrnove in the U band and Figure \ref{1999acmagutime} reports the
LC as well our fit.
A fourth   example  is   \snrel  (type Ia), which   has  been
analyzed  in \cite{Krisciunas2003}. Figure
\ref{2001elmagvtutto} reports the decay of the $V$   magnitude of
\snrel. A  comparison should be made with  Figure 4  of
\cite{Kasen2006} which uses the decay  of $^{56}$Ni.

\begin{figure}
\begin{center}
\includegraphics[width=6cm]{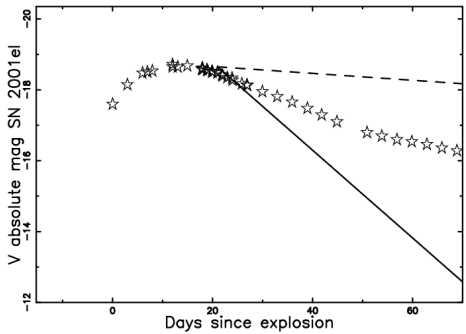}
\end {center}
\caption {
The $V$
LC  of \snrel (empty stars)
in absolute magnitude,
theoretical curve as given
by Equation (\ref{mgreatstandard})
when
the radioactive decay of the  isotope $^{56}$Ni
($\tau $ = 8.757 d  or  $T_{1/2}$ = 6.07 d)
was considered
(full line),
and
theoretical curve
of
the radioactive decay of the  isotope $^{56}$Co
($\tau $ = 111.47 d  or  $T_{1/2}$ = 77.27  d)
was considered
(dashed line).
}
\label{2001elmagvnuclear}
    \end{figure}

It is also interesting  to plot  the decay
of the LC  of  \snrel,
 see  \cite{Krisciunas2003},
as  given  by two nuclear  decay
which, according to
Equation (\ref{mgreatstandard}),  are straight lines,
see  Figure \ref{2001elmagvnuclear}.

\section{Conclusions}

The SN's are classified as spherical SN,
as an example \snr,
and as aspherical SN, see as  an example \cite{Racusin2009}
for \sn1987a.
The theory here developed treats the spherical
SN using classical
dimensional arguments.
The conversion of the flux of kinetic energy into
luminosity after
the maximum in the LC  explains the   curve of SNs  in
a direct form,
see  Equations (\ref{kineticflux}) and  (\ref{kineticfluxastro})
as  well as in a  logarithmic version,
see Equation (\ref{defmagnostra}).
The overall  LC  before and after
the maximum  can be built by introducing
two different physical
regimes, see  Equation (\ref{piecewise}).
The initial  rise in
intensity in the V-band is characterized
by a typical time scale
of $t_a \approx 5$ days  and the decrease
can be theoretically
fitted for $t \approx 3500$ days.
This  large  range in time  is also
the great advantage  of  our model:
the existing  nuclear models cover
$\approx$ 100 days, see  Figure 2 in  \cite{Leibundgut2003}.
The standard approach  of
formula (\ref{mstandard}), which predicts a linear increase  in the
apparent magnitude with time, does not  correspond  to the
observations because the observed and theoretical
magnitudes scale
as $m =a +b\, \ln(t)$ where $a$ and $b$ are two constants.
As  an example, Figure (\ref{2001elmagvnuclear})
reports two  commonly  accepted sources  which are
the radioactive  isotopes
$^{56}$Co, see  \cite{Georgii2000,Pluschke2001,Georgii2002},
 and
$^{56}$Ni,
see \cite{Truran2012,Dessart2012}:
the radioactive fit
is acceptable only  for the first few days.
The
application of the new formulas to three  SNs in different bands
gives acceptable results.
As an example, Figure  \ref {1993magtime}
reports the  LC  in the R-band for \snr  and Figure
\ref{1993halfatime} reports the LC for
the $H\alpha$ of
\snr.
An example of the two phase model as given by Equation
(\ref{piecewise}) is reported in Figure (\ref{2005magvtutto})
for
\snrcinque in the V-band.
A careful  analysis of the previous
figures shows that the theoretical and observed curves
present different concavities in the transition
from small to large times.
Similar  results can be obtained
assuming that all  $\gamma$-rays produced
by the decay of
$^{56}$Ni and
$^{56}$Co  are converted into optical emission,
see Figure 2 in  \cite{Leibundgut2003}.
The observational fact that the initial velocity
can be $\approx$ 30000 km s$^{-1}$
requires a relativistic treatment that
is necessary for future progress.
The analysis  here performed treats  the SN
as a single object
and therefore  is not connected with various types
of recent cosmologies, see
\cite{Astier2012,Chavanis2013,ElNabulsi2013}.

We conclude with a list  of not yet solved  problems:
\begin{itemize}
\item The observational fact that the initial velocity
can be $\approx 30000$ km s$^{-1}$
requires a relativistic treatment
of the flux of kinetic energy
that is
left for future research;
\item The connection between the cosmic ray production
and the $\gamma$-rays in SNR, see \cite{Dermer2013},
requires  an analysis of the temporal
behavior  of the magnetic field.
\end{itemize}


\end{document}